\documentclass[letterpaper, 10 pt, conference]{ieeeconf}  % Comment this line out if you need a4paper

\IEEEoverridecommandlockouts 
\usepackage{placeins} 
\usepackage{cite}
\usepackage{amsmath,amssymb,amsfonts}
\usepackage{algorithmic}
\usepackage{graphicx}
\usepackage{textcomp}
\usepackage{xcolor}
\usepackage{svg}
\usepackage{epsf,exscale,times}
\usepackage[english]{babel}
\usepackage{fancyhdr}
\usepackage{siunitx}
\usepackage{subcaption}
\usepackage{algorithm}
\usepackage[acronym]{glossaries}
\usepackage{hyperref}
\usepackage{ifthen}
\usepackage{textgreek}
\usepackage{ifthen}
\usepackage{gensymb}
\usepackage{tikz}
\usepackage{textcomp} % for the tradEMArk symbol
\usepackage{mathtools}
\usepackage{microtype}
\usepackage{booktabs}   % For \toprule, \midrule, \bottomrule
\usepackage{multirow}
\newacronym{6dof}{6-DoF}{six degrees of freedom}
\newacronym{MI}{MI}{Mutual information}
\newacronym{MINE}{MINE}{Mutual Information Neural Estimation}
\newacronym{GP}{GP}{Gaussian process}
\newacronym{MLAT}{MLAT}{multilateration}
\newacronym{TOA}{TOA}{time of arrival}
\newacronym{UWB}{UWB}{ultra wideband}
\newacronym{PMF}{PMF}{probability mass function}
\newacronym{UE}{UE}{user element}
\newacronym{CIR}{CIR}{channel impulse response}
\newacronym{PSO}{PSO}{particle swarm optimization}
\newacronym{LOS}{LOS}{line of sight}
\newacronym{SGA}{SGA}{stochastic gradient ascent}
\newacronym{BN}{BN}{Batch Normalization}
\newacronym{ELU}{ELU}{Exponential Linear Unit}
\newacronym{CPU}{CPU}{central processing unit}
\newacronym{GPU}{GPU}{graphics processing unit}
\newacronym{RAM}{RAM}{random access memory}
\newacronym{EMA}{EMA}{exponential moving average}
\newacronym{DOP}{DOP}{dilution of precision}
\newacronym{SVD}{SVD}{singular value decomposition}
\newacronym{CRB}{CRB}{Cramér–Rao bound}
\newacronym{FIM}{FIM}{Fisher information matrix}
\newacronym{TWR}{TWR}{two-way ranging}
\newacronym{RMSE}{RMSE}{root-mean-square error}
\newacronym{AMR}{AMR}{autonomous mobile robot}
\newacronym{RNN}{RNN}{recurrent neural network}
\newacronym{LSE}{LSE}{log sum exponent}
\newacronym{RSS}{RSS}{received signal strength}
\newacronym{MVUE}{MVUE}{Minimum-variance unbiased estimator}
\newacronym{MAE}{MAE}{mean absolute error}
\newacronym{GAE}{GAE}{geometric average error}
\newacronym{dFoV}{dFoV}{diagonal field-of-view}
\newacronym{FoV}{FoV}{field-of-view}
\newacronym{RANSAC}{RANSAC}{random sample consensus}
\newacronym{GNSS}{GNSS}{global navigation satellite system}
\newacronym{IPS}{IPS}{indoor positioning system}
\newacronym{IR}{IR}{infrared}
\newacronym{LiDAR}{LiDAR}{light detection and ranging}
\newacronym{radar}{radar}{radio detection and ranging}
\newacronym{IMU}{IMU}{inertial measurement unit}
\newacronym{RF}{RF}{radio frequency}
\newacronym{PnP}{PnP}{Perspective-n-Point}
\newacronym{EKF}{EKF}{Extended Kalman filter}
\newacronym{SLAM}{SLAM}{simultaneous localization and mapping}
\newacronym{ODOP}{ODOP}{orientational dilution of precision}
\newacronym{DLT}{DLT}{direct linear transform}
\newacronym{MCP}{MCP}{multiple coverage probability}
\newacronym{EGA}{EGA}{elimination genetic algorithm}
\newacronym{DWNA}{DWNA}{discrete white noise acceleration}
\newacronym{LCD}{LCD}{liquid-crystal display}
\newacronym{SURF}{SURF}{speeded up robust features}
\title{\LARGE \bf Investigation of ArUco Marker Placement for Planar Indoor Localization 
}

\author{
    Sven Hinderer$^{*}$, 
	Martina Scheffler$^{*}$, Bin Yang
    \thanks{$^{*}$Equal contribution.}%
	\thanks{All authors are with the Institute of Signal Processing and System Theory, University of Stuttgart, Stuttgart, Germany.
		{\tt\small firstname.lastname@iss.uni-stuttgart.de}}
}
\begin{document}
\maketitle
\thispagestyle{empty}
\pagestyle{empty}

\noindent\textbf{Note:} This is the accepted version of the paper. 
The final version is published in the \emph{Proceedings of the IEEE 2025 22nd International Conference on Advanced Robotics (ICAR)}. 
DOI: \href{https://doi.org/10.1109/ICAR65334.2025.11338671}{10.1109/ICAR65334.2025.11338671}
\vspace{1em}

\renewcommand{\thefootnote}{}
\footnotetext{
	\textcopyright~2025 Personal use of this material is permitted.  Permission from IEEE must be obtained for all other uses, in any current or future media, including reprinting/republishing this material for advertising or promotional purposes, creating new collective works, for resale or redistribution to servers or lists, or reuse of any copyrighted component of this work in other works
}
\renewcommand{\thefootnote}{\arabic{footnote}}

\begin{abstract}
	Indoor localization of \glspl{AMR} can be realized with fiducial markers. Such systems require only a simple, monocular camera as sensor and fiducial markers as passive, identifiable position references that can be printed on a piece of paper and distributed in the area of interest. Thus, fiducial marker systems can be scaled to large areas with a minor increase in system complexity and cost. We investigate the localization behavior of the fiducial marker framework ArUco w.r.t. the placement of the markers including the number of markers, their orientation w.r.t. the camera, and the camera-marker distance. In addition, we propose a simple Kalman filter with adaptive measurement noise variances for real-time \gls{AMR} tracking.
\end{abstract}
\vspace{1ex}

\section{Introduction}
Indoor localization \cite{review_indoor_loc} and mobile robotics \cite{ review_mobile_robots} have seen an increasing interest in the last years. As \gls{GNSS} cannot be used indoors, dedicated \glspl{IPS} have to
be deployed for indoor localization of \glspl{AMR}.

Indoor localization for \glspl{AMR} can be realized with a vast number of different technologies. These include camera, \gls{IR}, \gls{LiDAR}, \gls{radar}, (ultra)sound, Wi-Fi, Bluetooth, pseudolites, \gls{UWB}, magnetic localization, inertial navigation etc. An in-depth overview of these technologies can be found in \cite{review_indoor_pos_habil}.
Further, multiple modalities can be fused to compensate for weaknesses of one technology. To enable global localization and correct drift, \glspl{IMU} are e.g. often combined with other techniques.
Depending on the application, different requirements are posed on such systems. These include (among others) accuracy, cost, market maturity, update rates, robustness, and scalability \cite{review_indoor_pos_habil}.

Camera-based localization approaches can achieve cm localization accuracy at low system cost by making use of low-cost cameras (often monocular or even monochromatic) with high spatial resolution. Camera-based systems don't suffer from multi-path or interference, as e.g. \gls{RF} systems, but are affected by motion blur, illumination conditions, occlusion, accuracy degradation with distance, limited \gls{FoV}, high yet still limited resolution, and other hardware effects, camera settings etc. A survey of optical positioning systems is given in \cite{Survey_Optical_pos}. A newer overview of camera-based indoor localization approaches can be found in \cite{cv_loc_survey}, which classifies the different methods based on the environment data, sensing devices, detected elements, and the localization method. Adopting their classification, we study an environment with fiducial markers (landmarks) at known positions with a camera mounted on top of an \gls{AMR} for 2D camera localization. The detected elements are fiducial markers and localization is performed with traditional computer vision techniques. These markers can be printed on a piece of paper, are detectable under difficult conditions, and allow marker
identification by marker encoding. Such systems can therefore be scaled to large areas with only minor increase in system complexity. They have further shown to enable centimeter to sub-millimeter localization accuracy, achievable through high-end cameras and custom markers \cite{ODOP}, or small camera-marker distances \cite{QR_docking_submm}.

However, fiducial marker localization also has disadvantages \cite{Fiducial_Models} including camera pose ambiguity during frontal observation of the marker, dependence on marker size (marker must be covered in camera's \gls{FoV} while being large enough to provide sufficient information), degradation with imperfect camera calibration, and the previously mentioned camera system limitations.

In our localization system, we mount a camera facing the ceiling on top of an \gls{AMR}. On the ceiling, we distribute fiducial markers from the popular ArUco \cite{ArUco} framework. Each marker has four known 3D corner points in world coordinates and gives four detectable 2D corner points in the image. An image of multiple detected markers is shown in Fig.~\ref{fig:circ_meas}. Using these known corresponding pairs and a calibrated camera, the \gls{6dof} camera pose (rotation and translation) in world coordinates can be estimated with a \gls{PnP} algorithm, as explained in more detail in Sec.~\ref{sec:localization}. If more than one marker ($>4$ pairs) is used, we speak of multi-marker camera pose estimation. Otherwise we use single-marker (which is sufficient for \gls{6dof} camera pose estimation in case of planar markers like ArUco markers). Likewise, we denote multi-marker placement as fiducial marker placement for multi-marker camera pose estimation. The camera's $x$- and $y$- coordinates are then used as 2D \gls{AMR} coordinates for planar \gls{AMR} localization. 

\textbf{Contributions:}
In this work, we experimentally study the behavior of a popular fiducial marker framework ArUco \cite{ArUco} for planar indoor localization using the existing OpenCV framework \cite{opencv_library}. Our contributions are as follows:
\begin{itemize}
    \item We design and realize an indoor localization system based on the ArUco markers.
	\item We experimentally study the effect of multi-marker placement. To be more precise, we study the impact of the number of markers, their orientation w.r.t. the camera, and their distance to the camera on the localization accuracy.
    For fiducial marker systems, currently no viable studies about multi-marker placement exist.
	\item We propose a low-cost Kalman filter for real-time camera pose tracking that processes the \gls{PnP} camera pose estimates. Further, the Kalman filter adapts its measurement covariance matrix $\mathbf{R}^m$ based on the number of detected markers.
	
\end{itemize}

\section{Related work}

\subsection{Fiducial markers and ArUco}
There exists a large number of fiducial marker based systems. A comprehensive overview and timeline for fiducial markers can be found in \cite{phd_fiducial}. The most popular marker systems are the planar, square, black and white fiducial markers AprilTag \cite{april,apriltag2, april_improved} and ArUco \cite{ArUco, ArUco3}. A recent comparison between different planar fiducial markers is given in \cite{ArUco_Planar_markers_journal}, where ArUco shows superior accuracy and speed for robotics applications. In contrast, ArUco gives worse result in \cite{Fiducial_Journal}, but still competitive performance with similar markers (including AprilTag) at low computational cost. The authors of \cite{phd_fiducial} recommend ArUco for general applications due to its good software. Besides, ArUco's software is actively improved, e.g. w.r.t. processing time in ArUco3 \cite{ArUco3}. It is also equipped with \gls{SLAM} functionality. At first, offline camera localization and marker mapping was proposed in \cite{ArUco_mapping_localization}. This was later extended for online processing \cite{ArUco_SPM_SLAM}, inclusion of keypoints to the map \cite{Aruco_UcoSLAM}, and utilizing maps built with cameras different from those used during inference \cite{ArUco_ReSLAM}. In addition, parts of ArUco are available in OpenCV \cite{opencv_library}, whose implementation we use in this work. ArUco offers extensive capabilities and easy setup, making it a popular choice for robotics practitioners, and suitable for our study.

\subsection{Evaluation of fiducial marker placement}
Although there exists a large corpus of fiducial marker literature and many systems strive to improve camera pose estimation or detection rate, the effect of marker placement, especially multi-marker placement as required for robust localization, is still underexplored. 

\Gls{DOP}-like objectives, e.g. \gls{FIM} or variance based objectives, have been applied to fiducial marker systems. In \cite{Fiducial_Models}, analytical models for the variance of single-marker pose estimates are proposed based on the distance and angle between camera and marker. The adaptive, analytically computed variances are then used in \gls{EKF} localization, where a single-camera pose estimate serves as one input to the \gls{EKF}. While their adaptive variance models yield overall higher localization accuracy than static variances, this only holds in certain areas of the \gls{AMR} trajectory and multi-marker placement has not been investigated. The authors of \cite{ODOP} use precise binocular cameras and propose \gls{ODOP} to quantify the estimated rotation error. However, \gls{ODOP} is derived under strict pose error assumptions, which often don't hold due to the highly nonlinear camera pose estimation process with fiducial markers.

Experimental evaluations of fiducial markers w.r.t. different parameters can be found in many publications, including most fiducial marker systems described in \cite{phd_fiducial, Fiducial_Journal, ArUco_Planar_markers_journal}. In \cite{accuracy_artoolkit}, an accuracy function for single ARToolKit \cite{artoolkit} markers based on camera-marker distance and angle is defined. This was later extended by \cite{pentenrieder} with more simulated data. Similar experiments are found in \cite{accuracy_marker_3D} for AprilTags. This includes multi-marker placements, but only with simulated data and no in-depth investigation of the number of markers and multi-marker layout.
The authors of \cite{Fiducial_Journal} compare real measurements of single-, two- and three-marker placements. They include evaluations of accuracy and detection rate w.r.t. camera-marker distance, angle, illumination, motion blur, and two different webcams. They also give computational times. While containing some multi-marker placement experiments, they are restricted to a maximum of three fiducial markers and contain no investigation of the possible layouts of the multi-marker placement. \cite{ArUco_Planar_markers_journal} also compares different single, planar fiducial markers regarding their accuracy and detection rate based on camera-marker distance and angle, marker corner detection inaccuracies, occlusion, and computational times.

\subsection{Optimization of fiducial marker placement}
Fiducial marker placement optimization has been performed in  \cite{landmark_placement_navigation_cond_mi}. They optimize \gls{MI} of robot states along a full trajectory, given its observations. They extended this work in \cite{landmark_placement_navigation}. Here, 
landmarks are placed such that  a certain probability of maximum deviation from the ground truth for an \gls{AMR} trajectory is guaranteed with a minimum number of landmarks. They include ARToolKit as landmarks for evaluation. The authors of \cite{landmark_optimization_mcp}
introduce the concept of \gls{MCP}. This defines the probability of at least $n$ landmarks being visible by a camera at a fixed position. \gls{MCP} is computed based on camera resolution, \gls{FoV}, focus, occlusion, and camera orientation probability. Global marker placement optimization is then executed with an \gls{EGA} for simulated data. \cite{fiducial_optimization_entropy} adds a pre-defined set of markers to a scene such that the conditional entropy of the camera pose, given the measurements, is optimized with a greedy placement strategy. Opposed to other works, they optimize fiducial marker placements using both scene features (e.g. \gls{SURF} features from the environment) and fiducial markers. Their work is restricted to fiducial markers placed on a plane and multi-marker localization is not directly considered. They further optimize with simulated data and under Gaussian camera pose assumptions. While a Gaussian distribution is computationally attractive, allowing for an analytical entropy expression, it rarely applies for complex, fiducial based localization systems.

In contrast to previous work, we perform detailed evaluations of multi-marker placement for planar
\gls{AMR} localization. While multi-marker placement evaluations exist, they don't report accuracies w.r.t. the number of markers and often don't examine placement strategies applicable for planar \gls{AMR} localization. We collect real measurements and aim for fast processing, using the single camera position from \gls{PnP} camera pose estimation from multiple markers as input to the Kalman filter, opposed to the commonly applied, but more expensive \gls{EKF} localization \cite{prob_robotics}, fusing multiple single markers. This also deviates from the single \gls{EKF} pose input in \cite{Fiducial_Models} computed with a Gaussian product of individual marker estimates.

\section{Camera pose estimation with fiducial markers}
\label{sec:localization}
\subsection{Pinhole camera model}
For camera pose estimation, this work makes use of the pinhole camera model \cite[p.~153ff.]{Hartley_Zisserman_2004}, which assumes that an image is generated by rays of light passing through the camera's pinhole and hitting the sensor plane consisting of photo diodes. We define a 3D point in world coordinates as $\underline{p}_w=[x_w, y_w, z_w]^T\in\mathbb{R}^3$. The corresponding homogeneous coordinates are $\underline{\overline{p}}_w=[x_w, y_w, z_w, 1]^T\in\mathbb{R}^4$. Its 2D projection in image coordinates $\underline{p}_i=[x_i/z_i, y_i/z_i]^T\in\mathbb{R}^2$ can be computed from the homogeneous representation $\overline{\underline{p}
}_i=[x_i, y_i, z_i ]^T\in\mathbb{R}^3$. The projection of corresponding points in world coordinates $\underline{\overline{p}}_w$ to image coordinates $\underline{\overline{p}}_i$ is given by
\begin{align}
    \underline{\overline{p}}_i = \mathbf{P} \underline{\overline{p}}_w
\end{align}
with the transformation matrix $\mathbf{P}\in\mathbb{R}^{3\times4}$

\begin{equation}
            \overbrace{\mathbf{P}}^{\text{3$\times$4}}=
	        \overbrace{\begin{pmatrix}
			\mathbf{K}_{3\times3} \mid \underline{0}_{3\times1} \end{pmatrix}}^{\text{3$\times$4}} \cdot \overbrace{\left(\begin{array}{c|c}
		\mathbf{R}_{3\times3} & \underline{t}_{3\times1}\\
		\hline
		\underline{0}_{1\times3} & 1_{1\times1}
	\end{array}\right)}^{\text{4$\times$4}}.
\end{equation}

$\mathbf{P}$ contains two transformations. The transformation from a 3D point in the camera coordinate system to the corresponding 2D point on the image plane is described by the intrinsic matrix $\mathbf{K}\in\mathbb{R}^{3\times3}$ 
%\cite[p.~155]{Hartley_Zisserman_2004}
\begin{align}
\mathbf{K} = \begin{pmatrix}
    f_x & 0 & p_x \\
    0 & f_y & p_y \\
    0 & 0 & 1
\end{pmatrix},
\end{align}
where $f_x$ and $f_y$ are the focal lengths (the distance between the pinhole and the image plane) in $x$- and $y$- direction, and $[p_x, p_y]^T$ is the principal point (marking the intersection of the principal axis and the image plane).
$\mathbf{K}$ is typically estimated offline through camera calibration. The transformation from the world- to the camera coordinate system is expressed with the extrinsic parameters, namely the %\cite[p.~156]{Hartley_Zisserman_2004} 
rotation matrix $\mathbf{R}\in\mathbb{R}^{3\times3}$ and the translation vector $\underline{t}\in\mathbb{R}^{3}$. The goal for localization is then to estimate the extrinsic parameters and to extract the camera pose in world coordinates from them, as shown in \eqref{eq:camera_pose}.

\subsection{Perspective-n-Point problem}
For camera-based localization, the \gls{PnP} problem is the problem of estimating extrinsic parameters from known correspondences between $n$ 2D image points and $n$ 3D world points \cite[p.~178f.]{Hartley_Zisserman_2004}. Generally, the goal of the estimation is to find the 12 entries of the projection matrix $ \mathbf{P}$ \cite[p.~156]{Hartley_Zisserman_2004} to determine the \gls{6dof} camera pose (3 parameters for the translation vector and 3 independent rotations from the 3x3 rotation matrix) such that the reprojection error based on the Euclidean distances between the detected 2D image points $\underline{\hat{p}}_{ik}$ and the 2D image projections $\underline{p}_{ik}$ from corresponding 3D world points is minimized \cite[p.~181]{Hartley_Zisserman_2004}. From this, the 6 extrinsic parameters can be extracted if the intrinsic matrix $\mathbf{K}$ is known.
For a general solution of the \gls{PnP} problem, at least $n=6$ pairs $(\underline{p}_{ik}, \underline{\hat{p}}_{ik})$ are needed, though there exist solvers that are able to work with less points, if they fulfill additional criteria, allowing for \gls{6dof} camera pose estimation with a single marker. The solvers supported by OpenCV as well as their requirements can be found in the OpenCV documentation for \texttt{SolvePnPMethod} \cite{opencv_library}. For this work, the \texttt{ITERATIVE} scheme is used in combination with \gls{RANSAC} to gain more resistance to outliers. The solver itself is based on \gls{DLT} to find an initial estimate for $\mathbf{P}$ and a subsequent Levenberg-Marquardt optimization.

\subsection{Camera calibration}
The intrinsic matrix $\mathbf{K}$ is determined offline through camera calibration. During localization, $\mathbf{K}$ is passed to the OpenCV \gls{PnP} solver to obtain the extrinsic parameters.
Camera calibration requires at least ten images from different perspectives of a known calibration pattern. The images are then used to first estimate the homography for each view and then the intrinsic parameters \cite[p.~665f.]{kaehler_learning_2017}. 
Commonly used calibration patterns include a chessboard, ArUco or ChArUco patterns \cite[p.~652ff.]{kaehler_learning_2017}.
During this work, the chessboard pattern was used, as it led to the best pose estimates. Calibration with a \gls{LCD} instead of a printed version, as suggested in \cite{journal_lcd_calibration}, was examined and did not bring improvements in localization.

\subsection{Marker detection}
To solve the \gls{PnP} problem, correspondences between points in the image plane and in the world coordinate system are required. While the world points are measured during marker installation, the image points are detected with the provided ArUco software in OpenCV. After detecting the outer border of the marker, the inner pattern is extracted for determination of marker ID and rotation and the four corners of the quadratic marker are returned. For corner refinement, we apply the \texttt{CORNER REFINE CONTOUR} method
\cite{opencv_library} based on line fitting of detected contour points of the fiducial marker. These four image points are then used as detected 2D image plane points $\underline{\hat{p}}_{ik}$ in the \gls{PnP} solver.

\subsection{Random Sampling Consensus (RANSAC)}
Given a set of data points, the \gls{RANSAC} algorithm randomly selects a minimum number of them to solve for the (extrinsic) parameters. In a second step, the remaining set is divided into inliers and outliers, based on whether they fit the estimated model or not \cite[p.~117ff.]{Hartley_Zisserman_2004}. The OpenCV \texttt{solvePnPRansac} function \cite{opencv_library} makes use of this by randomly selecting a minimum number of correspondences, calling a \gls{PnP} solver for the selected pairs, and computing the reprojection errors to divide all samples into inliers and outliers. After repeating the procedure of random sampling, model fitting, and inlier/outlier division for a given number of iterations, \gls{RANSAC} returns the model with the most inliers. \gls{RANSAC} can deal with a large number of outliers, thus improving the robustness of localization.

\subsection{Camera pose estimation}
Using the detected marker corners and their measured positions in the world coordinate system as well as the intrinsic parameters gained by camera calibration, the solver can be called to estimate $\mathbf{P}$ and directly extract $\mathbf{R}$ and $\underline{t}$.
Once the extrinsic parameters are extracted from $\mathbf{P}$, the \gls{6dof} camera pose (i.e. it's position $\underline{C}\in\mathbb{R}^3$ and the 3 Euler angles including the heading $\theta$, derived from the rotation matrix $\mathbf{R}_C\in\mathbb{R}^{3\times3}$) is computed with \cite[p.~163f.]{Hartley_Zisserman_2004}
\begin{align}
	\left(\begin{array}{c|c}
		\mathbf{R}_C & \underline{C}\\
		\hline
		\mathbf{0} & 1
	\end{array}\right)
    =
    \left(\begin{array}{c|c}
		\mathbf{R} & \underline{t}\\
		\hline
		\mathbf{0} & 1
	\end{array}\right)^{-1}.
    \label{eq:camera_pose}
\end{align}

\section{Kalman filter} \label{sec:kf}
The Kalman filter \cite{Kalman} is an efficient realization of the Bayes' filter \cite{prob_robotics}. Such a filter usually consists of two steps, namely a prediction and an update step, which are applied sequentially for tracking objects like \glspl{AMR}. In the prediction step, given all previous measurements $\underline{y}_{1:t-1}$ up to time $t-1$, the distribution $p(\underline{x}_{t} | \underline{y}_{1:t-1})$ of the object state $\underline{x}_t$  at time $t$ is predicted with the state transition model $p(\underline{x}_t | \underline{x}_{t-1})$ under a Markov process first order assumption and by marginalizing out all previous object states $\underline{x}_{t-1}$ using the Chapman-Kolmogorov equation \cite{prob_robotics}
\begin{align}
	p(\underline{x}_t | \underline{y}_{1:t-1}) &= \int p(\underline{x}_t | \underline{x}_{t-1}) \, p(\underline{x}_{t-1} | \underline{y}_{1:t-1}) \, d\underline{x}_{t-1}.
	\label{eq:bayes_pred}
\end{align}
The state is then updated by applying Bayes' rule and integrating the current measurement $\underline{y}_t$ with the likelihood $p(\underline{y}_t | \underline{x}_t)$
\begin{align}
	p(\underline{x}_t | \underline{y}_{1:t}) &= \frac{p(\underline{y}_t | \underline{x}_t) \, p(\underline{x}_t | \underline{y}_{1:t-1})}{\int p(\underline{y}_t | \underline{x}_t) \, p(\underline{x}_t | \underline{y}_{1:t-1}) \, d\underline{x}_t}.
\end{align}
Under the assumptions of linear process and measurement models and independent, additive, white Gaussian process and measurement noise, the Kalman filter realizes the prediction step with
\begin{align}
	\underline{\overline{x}}_{t} &= \mathbf{F} \underline{x}_{t-1} + \mathbf{B} \underline{u}_t \label{eq:kalman_pred}\\ 
	\mathbf{\overline{\Sigma}}_{t} &= \mathbf{F} \mathbf{\Sigma}_{t-1} \mathbf{F}^T + \mathbf{Q},
\end{align}
where $\mathbf{F}$ is the state transition matrix, $\mathbf{B}$ is the control matrix, $\underline{u}$ is a control input vector, $\mathbf{\Sigma}$ is the state covariance matrix, and $\mathbf{Q}$ is the process covariance matrix. Following the prediction, the state update is carried out using
\begin{align}
	\mathbf{K}^g_t &= \mathbf{\overline{\Sigma}}_{t} \mathbf{H}^T (\mathbf{H} \mathbf{\overline{\Sigma}}_{t} \mathbf{H}^T + \mathbf{R}^m)^{-1} \\
	\underline{x}_{t} &= \underline{\overline{x}}_{t} + \mathbf{K}^g_t (\underline{y}_t - \mathbf{H} \underline{\overline{x}}_{t}) \label{eq:kalman_update}\\
	\mathbf{\Sigma}_{t} &= (\mathbf{I} - \mathbf{K}^g_t \mathbf{H}) \mathbf{\overline{\Sigma}}_{t}.
\end{align}
Thereby, $\mathbf{K}^g$ is the Kalman gain, $\mathbf{H}$ is the measurement matrix, $\mathbf{R}^m$ is the measurement covariance matrix, $\underline{y}_t-\mathbf{H}\underline{\overline{x}}_{t}$ is the innovation, and $\mathbf{I}$ is an identity matrix of suitable dimension. Before the sequential procedure starts, the state (mean) $\underline{x}_0$ and the state covariance $\mathbf{\Sigma}_0$ are initialized.
\subsection{Kalman filter for fiducial marker based localization}
In our localization problem, the state $\underline{x}_t=\left[x_t, v_{x,t}, y_t, v_{y,t}\right]^T\in\mathbb{R}^4$ contains the 2D $x$- and $y$- coordinates of the \gls{AMR} (its height $z_0$ is assumed constant and known as common for \glspl{AMR}) and its velocities $v_{x}$ and $v_{y}$ at time $t$. We initialize the \gls{AMR} state with the first measured $x_{m,0}, y_{m,0}$ values and assume that the \gls{AMR} has zero velocity at the start, i.e. $v_{x,0} = v_{y,0} = 0$, $\underline{x}_0 = \left[x_{m,0}, 0, y_{m,0}, 0 \right]^T$. The state covariance matrix is initialized as $\mathbf{{\Sigma}}_{0}=\mathrm{diag}(\sigma^2_{x,0}, \sigma^2_{v_x,0}, \sigma^2_{y,0}, \sigma^2_{v_y,0})$, where $\sigma^2_{v_x,0}=\sigma^2_{v_y,0}=0$. The state variances for $x$ and $y$ are initialized with the first measurement variances $\sigma^2_{m,x}$ and $\sigma^2_{m,y}$ as described at the end of this section. 

The \gls{PnP} solver directly outputs the camera pose (including $x$ and $y$ used for \gls{AMR} localization), therefore the measurements $\underline{y}_t=\left[x_{m,t}, y_{m,t}\right]^T\in\mathbb{R}^2$ are the camera's $x$- and $y$- coordinates. The process matrix $\mathbf{F}\in\mathbb{R}^{4\times 4}$, assuming a constant \gls{AMR} velocity, is given by
\begin{equation}
	\mathbf{F} = \begin{pmatrix}
		1 & \Delta t & 0 & 0 \\
		0 & 1 & 0 & 0\\
		0 & 0 & 1 & \Delta t\\
		0 & 0 & 0 & 1
	\end{pmatrix}
	\label{eq:F}.
\end{equation}
Since we don't use a control input, the $\mathbf{B}\underline{u}_t$ term in \eqref{eq:kalman_pred} is omitted. The process noise matrix $\mathbf{Q}\in\mathbb{R}^{4\times 4}$ is defined as
\begin{equation}
	\mathbf{Q} = \begin{pmatrix}
		\mathbf{Q}_x & \mathbf{0} \\
		\mathbf{0} & \mathbf{Q}_y \\
	\end{pmatrix}
\end{equation}
with the noise components
\begin{align}
	\mathbf{Q}_{x} &= \begin{pmatrix}
		\frac{\Delta t^4}{4} & \frac{\Delta t^3}{2} \\
		\frac{\Delta t^3}{2} & \Delta t^2 \\
	\end{pmatrix} {\sigma^2_{a_x}} , \label{eq:sigma-v-x}\\
	\mathbf{Q}_{y} &= \begin{pmatrix}
		\frac{\Delta t^4}{4} & \frac{\Delta t^3}{2} \\
		\frac{\Delta t^3}{2} & \Delta t^2 \\
	\end{pmatrix} {\sigma^2_{a_y}}.\label{eq:sigma-v-y}
\end{align}
In this work, $\mathbf{Q}_x\in\mathbb{R}^{2\times 2}$ and $\mathbf{Q}_y\in\mathbb{R}^{2\times 2}$ are defined using
a \gls{DWNA} model, which assumes that the constant velocity system is
disturbed by a piecewise constant acceleration,
where the accelerations per time step are uncorrelated \cite[p.~273]{bar-shalom_estimation_2002}. The variances for the accelerations in $x$ and $y$ are $\sigma^2_{a_x}$ and $\sigma^2_{a_y}$, respectively. We experimentally set $\sigma^2_{a_x}=\sigma^2_{a_y}=1\left(\frac{\mathrm{cm}}{\mathrm{s}^2}\right)^2$ such that measurements are trusted more than the motion model.

The measurement matrix $\mathbf{H}\in\mathbb{R}^{2\times 4}$ is
\begin{equation}
	\mathbf{H} = \begin{pmatrix}
		1 & 0 & 0 & 0\\
		0 & 0 & 1 & 0\\
	\end{pmatrix}
	\label{eq:H}
\end{equation}
and the measurement covariance matrix $\mathbf{R}^m\in\mathbb{R}^{2\times 2}$ is given by $\mathbf{R}^m=\mathrm{diag}(\sigma^2_{m,x}/4, \sigma^2_{m,y}/4)$ with factor $1/4$ for improved tracking results.
The measurement variances are determined experimentally with~Fig.~\ref{fig:circ_meas} for different numbers of detected markers and shown in Table~\ref{tab:variance_vs_detections}.
\begin{table}[t]
    \centering
    \caption{Measured variances $\sigma^2_{m,x}$ and $\sigma^2_{m,y}$ based on the number of detected markers.}
    
    \label{tab:variance_vs_detections}
    \begin{tabular}{c | r r r}
        \toprule
        \# Detections & $\sigma^2_{m,x}$ (cm$^2$) & $\sigma^2_{m,y}$ (cm$^2$) \\
        \midrule
        1 & 1944.56 & 914.43 \\
        2 & 603.46 & 464.10\\
        3 & 68.88 & 64.47\\
        4 & 19.35 & 16.77\\
        5 & 7.62 & 6.64\\
        6 & 4.46 & 3.49\\
        7 & 3.04 & 2.09\\
        \bottomrule
    \end{tabular}
\end{table}
Our Kalman filter therefore adapts $\mathbf{R}^m$ based on the number of detected markers.
\section{Experiments}
\subsection{Experimental environment}
Our experiments are carried out in a small laboratory equipped with a chessboard floor with known grid size of ($16.\overline{6}$, $16,\overline{6}$) cm. The known grid coordinates are used as position references to determine the camera's and marker's $x$- and $y$- coordinates and orientation $\theta$. To accurately install the hardware, we use a laser rangefinder to determine the height of the camera or markers. Markers are installed either on the floor or on the ceiling. For ceiling installations, we project the known $x$- and $y$- coordinates on the floor to the ceiling with a laser level. The camera is a Logitech C920 Pro HD with a resolution of 1080p and a \gls{dFoV} of $78^\circ$. The markers are created with an ArUco $6 \times 6$ dictionary, which provides 1000 different markers (\texttt{DICT\_6X6\_1000}), and printed with a square marker size of (15, 15) cm.
\subsection{Camera and marker configurations}
\begin{figure}[h!]
	\centering
	\begin{subfigure}[b]{.47\textwidth}\
    \hspace*{0.15\textwidth}
	\includesvg[width=0.7\textwidth]
		{circle_sketch.svg}
		\caption{Circular marker placement including marker IDs and coordinate system.}
		\label{fig:circ_layout}
	\end{subfigure}
    
	\vfill
    
	\begin{subfigure}[b]{.47\textwidth}
		\includegraphics[width=\textwidth]{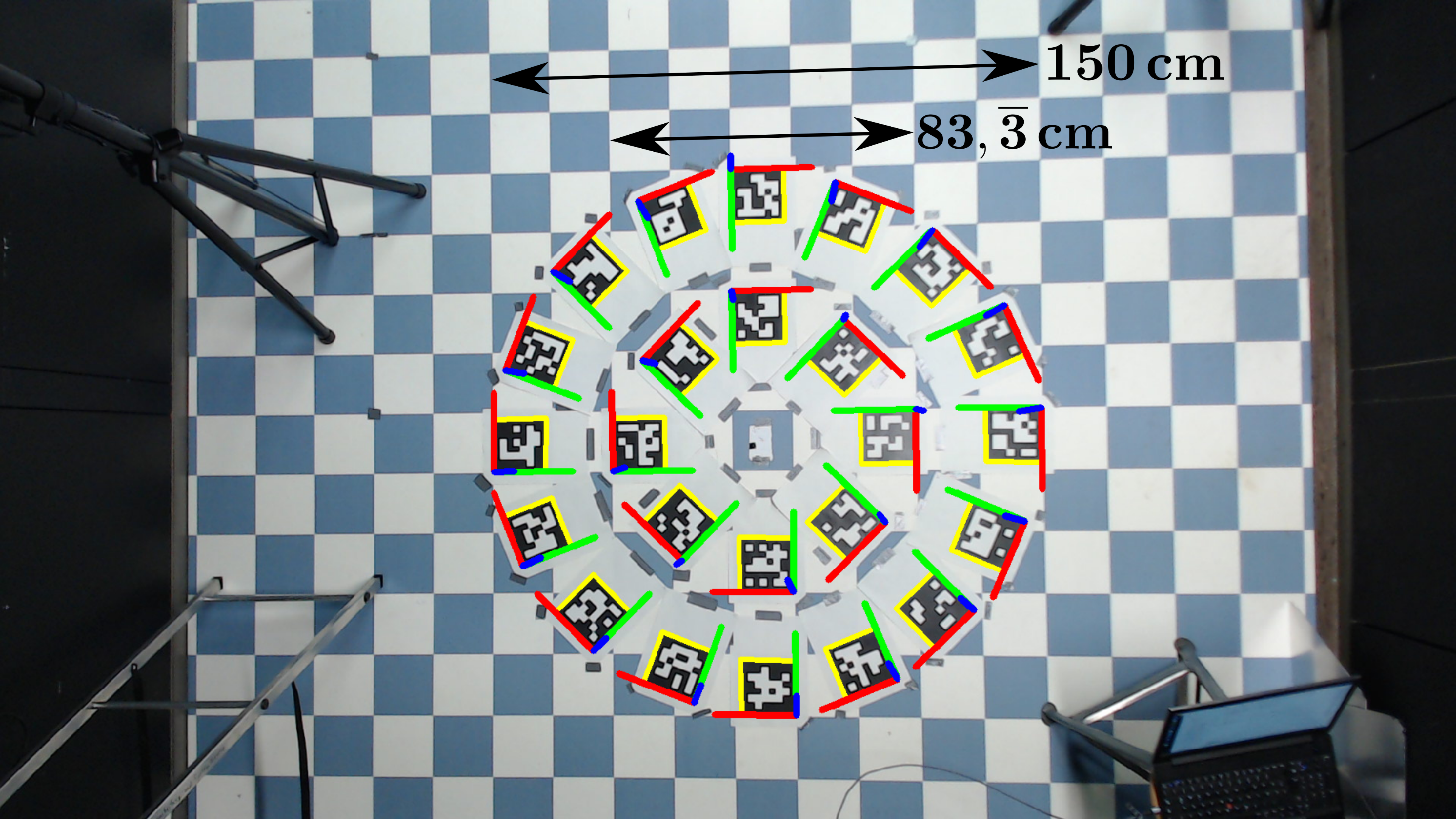}
		\caption{Respective marker detections in image collected at a camera height of \SI{2.991}{m}.}
		\label{fig:circ_meas}
	\end{subfigure}
    
	\caption{In a) is our marker placement for marker placement evaluation, where markers form two circles (inner: markers 0-7, outer: markers 8-23) on the ground. Three sets of marker rotations (w.r.t. the closest chessboard lines in angle) are defined, where markers with IDs from $\{0, 8, 2, 12, 4, 16, 6, 20\}$ belong to the $90^\circ$ rotations, $\{1, 10, 3, 14, 5, 18, 7,22\}$ belong to the $45^\circ$ rotations, and $\{9, 11, 13, 15, 17, 19, 21, 23\}$ belong to the $22.5^\circ$ rotations. For real image collection, the camera is installed at a height of \SI{1.334}{m} or \SI{2.991}{m} (as in b)). The image collected at \SI{1.334}{m} misses a few markers due to limited camera \gls{FoV}.}
	\label{fig:circles}
\end{figure}

\textbf{Configuration I:} To simplify our marker placement, we place the markers on the floor and tape the camera to the ceiling. The markers form two circles, as shown in Fig.~\ref{fig:circles}.
This allows us to investigate different marker orientations, combinations, distances to the image center, and by placing the camera at different heights of $z=$ \SI{1.334}{m} and $z=$ \SI{2.991}{m}, also different camera-marker distances. 

\begin{figure}[h!]
	\centering
    \includesvg[width=.30\textwidth]
		{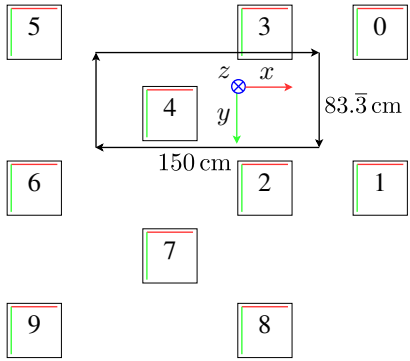}
		\caption{Marker placement (on ceiling) and rectangular \gls{AMR} movement (black arrows) for the tracking experiments.}

	\label{fig:kalman_layout}
\end{figure}

\textbf{Configuration II:} For the camera pose tracking experiments with our Kalman filter, the markers are placed on the ceiling and the camera is on the floor, as in the intended \gls{AMR} application. The markers are placed such that at least five markers are visible from each \gls{AMR} position (for high robustness as explained in the results), markers are spread out to increase their geometric aperture, and markers are not too close to lights on the ceiling. The multi-marker placement and the rectangular \gls{AMR} trajectory are depicted in Fig.~\ref{fig:kalman_layout}. To showcase the advantage of using the adaptive measurement variances, we use a maximum of 1, 3, 5 and all visible markers along the four linear trajectories. We then compare the adaptive variances with the static max/min/mean/median variances from Table~\ref{tab:variance_vs_detections} for $x$ and $y$, respectively, similar to the adaptive single-marker variance model evaluation in \cite{Fiducial_Models}.

\section{Results}
\subsection{Influence of camera-marker distance, marker orientation, and distance from image center}
\begin{figure}[h!]
    \centering
    \begin{subfigure}[b]{0.47\textwidth}
        \centering
        \scalebox{0.9}{\input{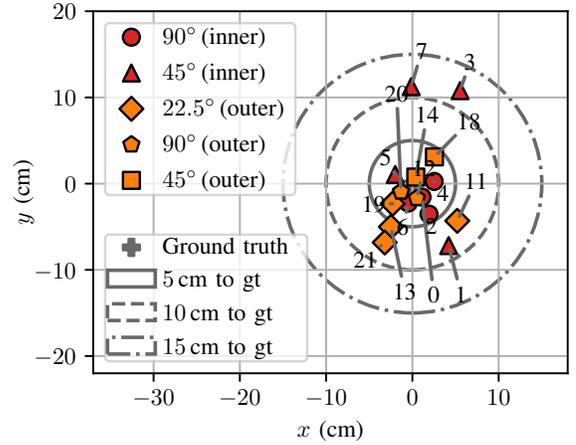}} 
        \caption{Camera height of \SI{1.334}{m}.}
        \label{fig:placement_eval_low}
    \end{subfigure}
    \hfill
    \begin{subfigure}[b]{0.47\textwidth}
        \centering
        \scalebox{0.9}{\input{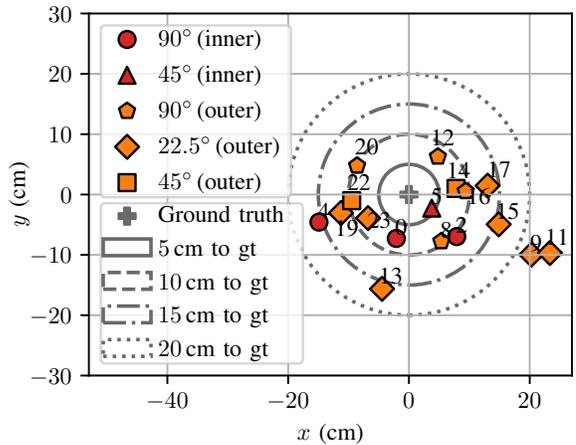}} % scale as needed
        \caption{Camera height of \SI{2.991}{m}.}
        \label{fig:placement_eval_high}
    \end{subfigure}
    \caption{Error distribution for the circular marker experiments in configuration I with camera heights of a) \SI{1.334}{m} and b) \SI{2.991}{m}. Fiducial markers are encoded with the definitions from Fig.~\ref{fig:circles}.}
    \label{fig:placement_eval}
\end{figure}

The two different camera-marker distances in configuration I show the expected increase in localization error with larger camera-marker distances, as evident when comparing Table~\ref{tab:placement_eval_low} with Table~\ref{tab:placement_eval_high}, and Fig.~\ref{fig:placement_eval_low} with Fig.~\ref{fig:placement_eval_high}. Fig.~\ref{fig:placement_eval_high} misses outliers. It can be observed in both figures and tables that inner markers perform worse than the outer markers. In the experiments conducted for Fig.~\ref{fig:placement_eval_high}, 50\% of the markers on the inner circle show errors greater than $20\,\text{cm}$ (with many outliers not included in the figure), whereas for the outer circle it's only 31.25\%. This suggests that markers placed further from the image center are more beneficial for localization, likely because inner markers are closer to the problematic frontal marker observation. In additional experiments with two-marker placements, distributing them over the image space and placing them far from the center also gave the best results, as the markers cover a larger geometric aperture. 

Regarding fiducial marker orientation, Table~\ref{tab:placement_eval_low} and Table~\ref{tab:placement_eval_high} show the best performance for the 90$^{\circ}$ type markers, which we suspect to be caused by better marker detection with OpenCV and more precise marker installation for this type. Different reprojection error weights (e.g. based on camera-marker distance and angle) in the \gls{PnP} solver might be studied in future work to enhance multi-marker localization.

\begin{table}[h!]
    \centering
    \caption{\Gls{RMSE} and mean error (bias) for the circular layout in configuration I with a camera height of \SI{1.344}{m}.}
    \label{tab:placement_eval_low}
    \begin{tabular}{l | r r r | r r r}
        \toprule
        \multirow{2}{*}{Marker type} & \multicolumn{3}{c|}{RMSE} & \multicolumn{3}{c}{Mean error} \\
        & $x$ (cm) & $y$ (cm) & $\theta$ ($^{\circ}$) & $x$ (cm) & $y$ (cm) & $\theta$ ($^{\circ}$) \\
        \midrule
        All                   & 2.75 & 5.16 & 0.36 & 0.75  & \textbf{-0.52} & -0.13 \\
        $90^{\circ}$ (inner)  & 1.70 & 2.24 & 0.28 & 1.27  & -1.77 & \textbf{0.01} \\
        $90^{\circ}$ (outer)  & \textbf{1.00} & \textbf{1.37} & 0.77 & \textbf{-0.41} & -1.31 & -0.42 \\
        $45^{\circ}$ (inner)  & 3.61 & 8.60 & \textbf{0.01} & 1.90  & 3.98  & 0.24 \\
        $45^{\circ}$ (outer)  & 1.83 & 2.26 & 0.19 & 1.48  & 1.93  & -0.16 \\
        $22.5^{\circ}$ (outer)& 3.49 & 4.89 & 0.26 & -0.70 & -4.62 & -0.23 \\
        \bottomrule
    \end{tabular}
\end{table}
\begin{table}[h!]
    \centering
    \caption{\gls{RMSE} and mean error for the circular layout with a camera height of \SI{2.991}{m}.}   \label{tab:placement_eval_high}
    \begin{tabular}{l | r r r | r r r}
        \toprule
        \multirow{2}{*}{Marker type} & \multicolumn{3}{c|}{RMSE} & \multicolumn{3}{c}{Mean error} \\
        & $x$ (cm) & $y$ (cm) & $\theta$ ($^{\circ}$) & $x$ (cm) & $y$ (cm) & $\theta$ ($^{\circ}$) \\
        \midrule
        All                   & 45.29 & 31.38 & 1.04 & 10.34  & -8.38 & -0.03 \\
        $90^{\circ}$ (inner)  & 36.97 & 5.66  & \textbf{0.42} & 15.72  & -5.30 & \textbf{0.01} \\
        $90^{\circ}$ (outer)  & \textbf{7.31} & \textbf{5.53} & 0.52 & 2.74 & \textbf{0.95} & -0.25 \\
        $45^{\circ}$ (inner)  & 46.52 & 51.10 & 1.87 & \textbf{1.43} & -21.10 & 0.85 \\
        $45^{\circ}$ (outer)  & 61.86 & 53.22 & 1.45 & -3.67 & -5.00 & -1.17 \\
        $22.5^{\circ}$ (outer)& 49.49 & 14.27 & 0.49 & 22.93  & -9.94 & 0.19 \\
    \bottomrule
    \end{tabular}
\end{table}
\subsection{Influence of number of markers}
\begin{figure}[h!]
	\centering

    \hspace*{-0.02\textwidth}
	\input{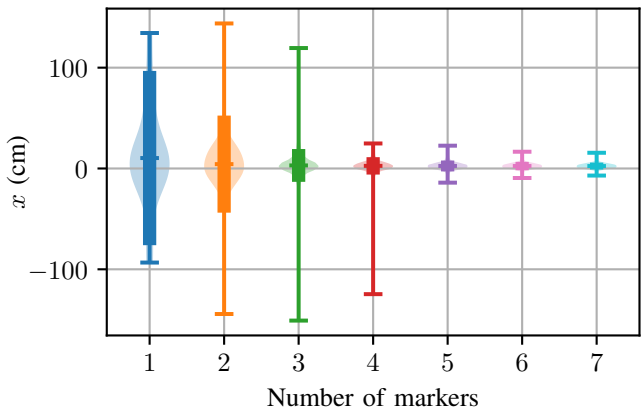}

	\caption{Violin plot of the errors in $x$ for different numbers of markers and a camera height of \SI{2.991}{m}.}
	\label{fig:num_markers_violin}
\end{figure}

\begin{figure}[h!]
    \centering
    \begin{subfigure}[b]{0.47\textwidth}
        \centering
        \scalebox{0.95}{\input{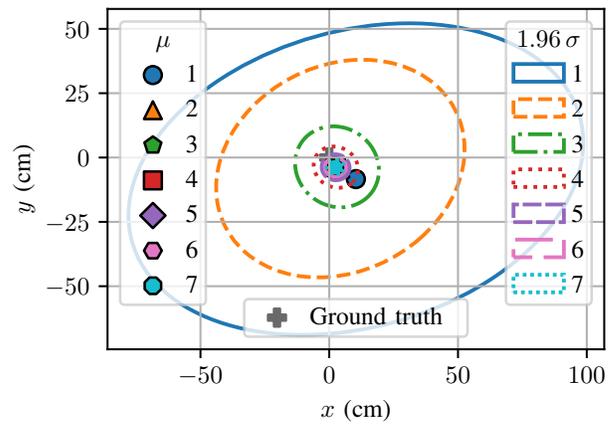}} % adjust scale as needed
        \caption{All evaluated numbers of markers.}
        \label{fig:num_markers_violin_all}
    \end{subfigure}
    \hfill
    \begin{subfigure}[b]{0.47\textwidth}
        \centering
        \scalebox{0.95}{\input{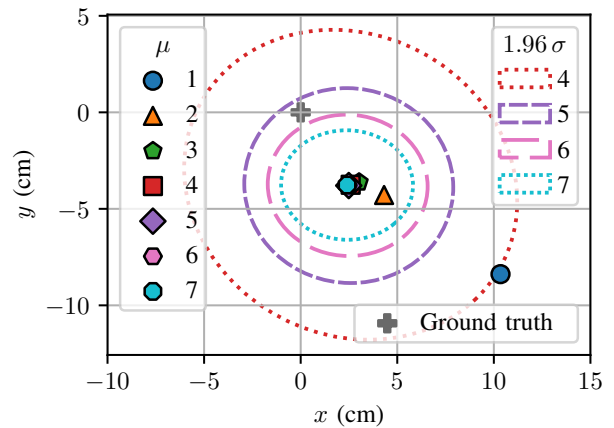}} % adjust scale as needed
        \caption{Zoom into region with 4+ markers.}
        \label{fig:num_markers_violin_zoom}
    \end{subfigure}
    \caption{Mean errors and confidence ellipses for a) all different numbers of markers and b) zoom into region with 4+ markers. The camera is installed at a height of \SI{2.991}{m}.}
    \label{fig:num_markers_ellipse}
\end{figure}

For the marker placement in Fig.~\ref{fig:circles}, all possible single- and multi-marker placement permutations with one to seven markers were analyzed. The errors in localization are shown in Table~\ref{tab:nr_markers} and their distributions are visualized in Fig.~\ref{fig:num_markers_violin} and Fig.~\ref{fig:num_markers_ellipse}. It can be seen that larger numbers of markers lead to better mean values and tighter distributions. However, using more than five markers leads to less improvements in localization, as can be seen in Table~\ref{tab:nr_markers} and the violin plots for $x$ (and similarly for $y$ and $\theta$, skipped due to limited space) in Fig.~\ref{fig:num_markers_violin}. At 5+ markers, there is a significant drop in maximum error, indicating a possible lower bound on the required number of visible markers for robust localization. It can also be seen that while more markers give lower \gls{RMSE}, maximum errors can be larger with more markers. Further, we notice a systematic error in Table~\ref{tab:nr_markers} and Fig.~\ref{fig:num_markers_ellipse} that does not decrease by adding markers, caused by imperfect camera calibration and possibly other static error sources such as installation errors for camera- and marker positions, camera hardware/settings etc.
\begin{table}[t]
    \centering
    \caption{\gls{RMSE} and mean error for different numbers of markers with a camera height of \SI{2.991}{\meter}.}
    \label{tab:rmse_mean_markers}
    \begin{tabular}{c | r r r | r r r}
        \toprule
        \multirow{2}{*}{Nr. of markers} & \multicolumn{3}{c|}{RMSE} & \multicolumn{3}{c}{Mean error} \\
        & $x$ (cm) & $y$ (cm) & $\theta$ ($^{\circ}$) & $x$ (cm) & $y$ (cm) & $\theta$ ($^{\circ}$) \\
        \midrule
        1 & 45.29 & 31.38 & 1.04 & 10.34 & -8.38 & -0.03 \\
        2 & 24.94 & 21.96 & 0.59 & 4.32 & -4.28 & -0.09 \\
        3 & 8.83  & 8.81  & 0.16 & 3.02 & -3.62 & -0.06 \\
        4 & 5.10  & 5.55  & 0.10 & 2.59 & -3.75 & -0.06 \\
        5 & 3.72  & 4.59  & 0.08 & 2.49 & -3.79 & -0.07 \\
        6 & 3.22  & 4.22  & 0.07 & 2.43 & -3.78 & -0.07 \\
        7 & 2.97  & 4.03  & 0.07 & 2.40 & -3.77 & -0.07 \\
        \bottomrule
    \end{tabular}
    \label{tab:nr_markers}
\end{table}


\subsection{Kalman filter tracking}
\begin{figure}[h]
	\centering
	
	    \begin{subfigure}[b]{0.235\textwidth}
			\centering
		    \includegraphics[width=1.0\linewidth]{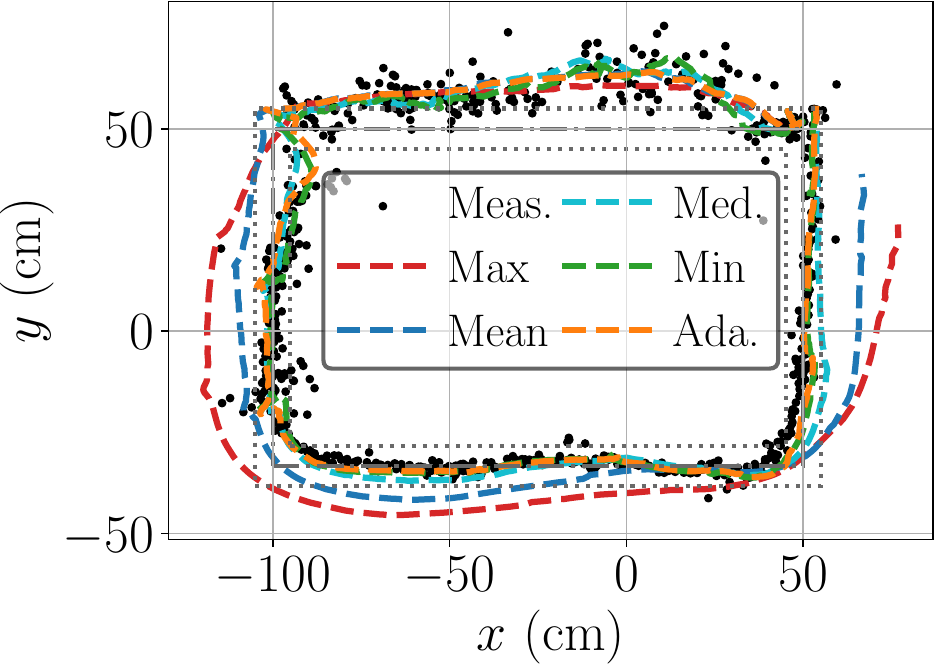}
		
			\caption{Best tracking performance with min and adaptive (Ada.) variance models.}
			\label{fig:kalman1}
		\end{subfigure}
	\hfill
	    \begin{subfigure}[b]{0.235\textwidth}
			\centering
			\includegraphics[width=1.0\linewidth]{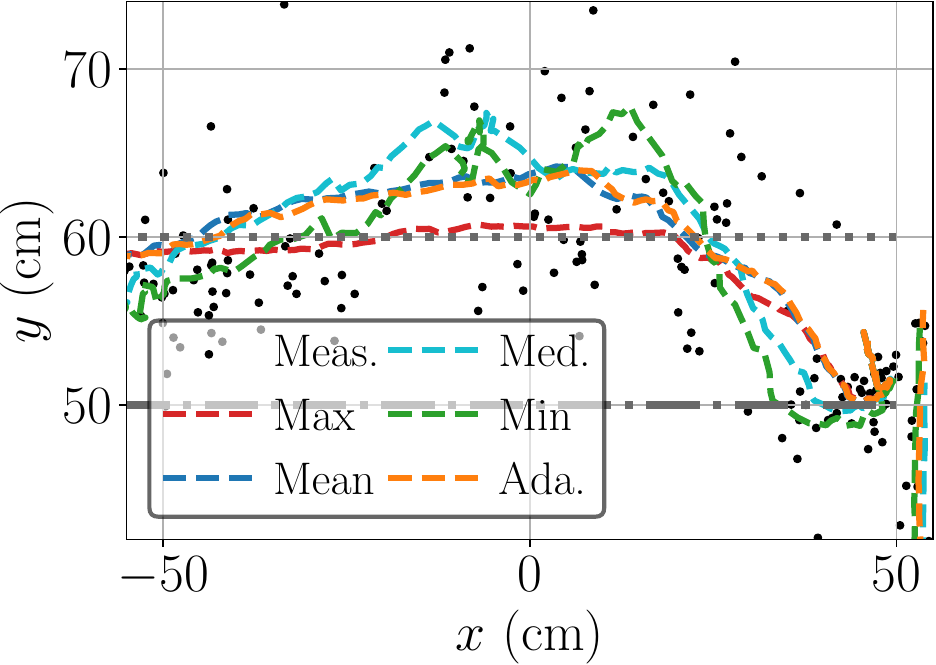}
			
			\caption{Adaptive variance model is more robust than min variance model.}
			\label{fig:kalman2}
	\end{subfigure}
			\caption{Kalman filter tracking results for a) the rectangular trajectory and b) the start region with single-marker positioning. Black dots are the estimated camera position inputs to the Kalman filter. The gray rectangles denote the hand-measured ground truth with 5 cm confidence interval.}
	\label{fig:kalman_result}
\end{figure}

For the experiments with our proposed Kalman filter in configuration II, the markers are placed on the ceiling and the camera is moved (by hand) on the floor as shown in Fig.~\ref{fig:kalman_layout}. The resulting ground truth is visualized in Fig.~\ref{fig:kalman_result} together with the estimated camera positions from camera pose estimation with measured fiducial markers and the posteriors $\underline{x}_t$ from \eqref{eq:kalman_update} for the different measurement variance models.
The max, mean and median variances are too large and cause notable deviations from the ground truth, as seen in Fig.~\ref{fig:kalman1}. The min variances lead to accurate \gls{AMR} trajectories if enough markers are visible (bottom with $\leq 5$ and right with all visible markers). However, when less markers are visible (left $\leq 3$ and top $\leq 1$), the min variances cause nonsmooth trajectories as shown in Fig.~\ref{fig:kalman2}. There, the Kalman filter becomes overconfident in the less reliable measurements. Our proposed adaptive measurement variances give high accuracy and smooth \gls{AMR} motion in all regions and overall best performance.

\section{Conclusion}
We studied the effect of ArUco marker placement on the accuracy of planar, camera based localization. Therefore, markers were placed in circles and single-marker placements were evaluated individually, and multi-marker placements  exhaustively. 

Results have shown clear localization improvements with smaller camera-marker distance. Markers further from the image center and $90^\circ$ markers with marker borders parallel to the chessboard floor lines (and parallel/orthogonal to the camera orientation) lead to slightly lower positioning errors. Some systematic error and asymmetry in $x$- and $y$- position estimation was observed, likely due to remaining calibration errors. Rotation errors are overall very low.

Multi-marker placements gave clear positioning accuracy improvements by using more markers for \gls{PnP} based camera pose estimation. With 5+ markers, the maximum positioning errors are much lower than with less than 5 available markers. The maximum error and large error quantiles are often more important for localization than the mean positioning error. With 5+ markers, reliable localization, even with the worst multi-marker layouts, is possible. However, with more than 5 markers, the gain in localization accuracy drops significantly. 

Further, a low-cost Kalman filter was introduced, that processes one single \gls{PnP} camera pose estimate per frame. This allows for fast camera and \gls{AMR} position estimation. 

Our work has several limitations. The evaluated area is small and good marker coverage with small camera-marker distances is always guaranteed. We only tried one fiducial marker framework with one camera. The linear Kalman filter relies on the measurements to update the \gls{AMR} velocity and performance will degrade with nonlinear motions.
Future work could perform more experiments by benchmarking the multi-marker behavior of different fiducial marker systems and in larger environments, including 3D placement (with markers on ceiling/floor/walls). Based on our results, we can conclude that markers should be placed such that 5+ markers must be visible from every feasible \gls{AMR} position if high reliability is required. 

Deeper investigation into optimizable objectives to quantify and optimize the localization accuracy with multi-marker placement should be done for better guidance of placement optimization algorithms. 
Finally, the tracking algorithm might be enhanced by integration of odometry, nonlinear tracking methods, and possibly multi-camera setups as in \cite{rss_gt_fiducial}.

\bibliographystyle{ieeetr}
\bibliography{ref}
\end{document}